\documentclass[aps,pra,amssymb,amsmath,twocolumn,superscriptaddress,showpacs]{revtex4-1}

\usepackage{graphicx}
\usepackage{bm}
\usepackage{amsfonts}
\usepackage{color}

\newcommand{\ket}[1]{\left|#1\right\rangle}
\newcommand{\bra}[1]{\left\langle #1\right|}

\begin{document}

\title{Experimental orbital angular momentum based \\ quantum key distribution through turbulence}

\author{Sandeep Goyal} 
\email{sandeep.goyal@ucalgary.ca}
\affiliation{Institute of Quantum Science and Technology, University of Calgary, Alberta T2N 1N4, Canada}  
\affiliation{School of Chemistry and Physics, University of Kwazulu-Natal, Private Bag X54001, Durban 4000, South Africa}
\author{Alpha Hamadou Ibrahim}
\affiliation{CSIR National Laser Centre, PO Box 395, Pretoria 0001, South Africa}
\affiliation{Department of Physics, University of South Africa, PO Box 392, Pretoria 0003, South Africa}
\author{Filippus S. Roux}
\email{fsroux@csir.co.za}
\affiliation{CSIR National Laser Centre, PO Box 395, Pretoria 0001, South Africa}
\author{Thomas Konrad} 
\affiliation{School of Chemistry and Physics, University of Kwazulu-Natal, Private Bag X54001, Durban 4000, South Africa}
\affiliation{National Institute of Theoretical Physics, University of Kwazulu-Natal, Private Bag X54001, Durban 4000, South Africa}
\author{Andrew Forbes} 
\affiliation{CSIR National Laser Centre, PO Box 395, Pretoria 0001, South Africa}
\affiliation{School of Physics, University of the Witwatersrand, Private Bag X3, Johannesburg 2030, South Africa}


\begin{abstract}
Using an experimental setup that simulates a turbulent atmosphere, we study the secret key rate for quantum key distribution protocols in orbital angular momentum based free space quantum communication. The quantum key distribution protocols under consideration include the Ekert 91 protocol for different choices of mutually unbiased bases and the six-state protocol. We find that the secret key rate of these protocols decay to zero roughly at the same scale where the entanglement of formation decays to zero.
\end{abstract}


\maketitle

\section{Introduction}

Quantum key distribution (QKD) is the first cryptographic method that is based on the laws of quantum mechanics. In principle, it provides a means to communicate securely against eavesdropping, by establishing a secret key between two authenticated parties that can be used for secret communication. This is the first experimentally realizable application of quantum information processing and has drawn the attention of a large community in both theory \cite{Lo1998,*Ekert2001,*Gisin2002, *Scarani2006,*Le2006,*Dusek2006,*Lo2009,*Scarani2009} and experiment \cite{groblacher2006,*Ursin2007etal,*Erven2008,*Peev2009etal,*Peloso2009,*Jin2010, *Sasaki2011etal,*Erven2012}.

To replace classical cryptographic technology, which are in general not secure against attacks using nascent quantum computing technology, methods are sought to achieve QKD with high transmission rates over large distances. Even though the implementation technique for QKD has reached the commercial level \footnote{Quintessence~Labs~(www.quintessencelabs.com), MagiQ Technologies~(www.magiqtech.com), idQuantique~(www.idquantique.com), and SeQureNet~(www.sequrenet.fr)}, the transmission distances and rates, are still comparatively small. The most robust quantum channels for QKD are currently based on optical fibers, with transmission distances of between 20~km and 150~km and maximal bit rates of between 10~kbit and 1~Mbit per second \cite{Dixon2008}.

\begin{figure}[ht]
\includegraphics{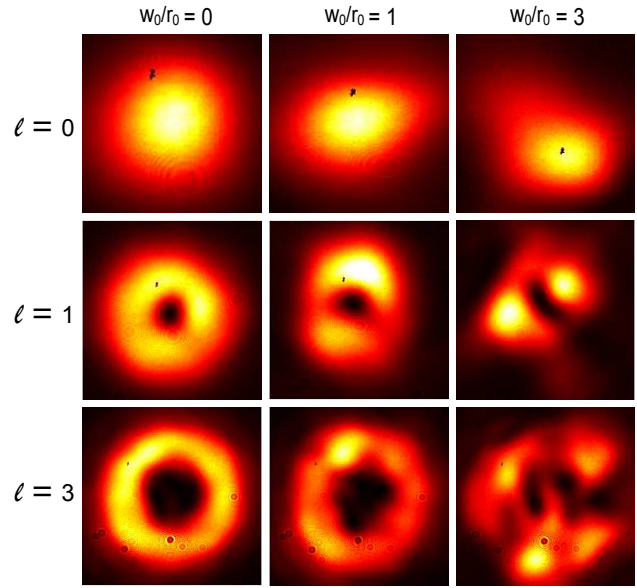}
\caption{(Color online) Experimentally measured effect of turbulence on the intensity distributions of OAM modes for $\ell = 0,1,3$. First column without turbulence (${\cal W}=0$). Second and third columns with progressively more severe turbulence conditions (${\cal W} = 1$ and 3, respectively). }
\label{beamdist}
\end{figure}

The free space implementation of QKD could enable intercontinental transmission channels using satellites \cite{boone2014}. A possible candidate as information carrier with high information capacity is the orbital angular momentum (OAM) modal basis of photons \cite{Allen1992}. However, while the existence of infinitely many OAM modes, in principle, allows one to encode an arbitrary amount of information in a single photon, these modes are susceptible to the influence of atmospheric turbulence (see Fig.~\ref{beamdist}). Apart from its adverse affect on spatial modal multiplexing in free space optical communication \cite{malik2012, *BRodenburg2012a, *rodenburg2014a, *rodenburg2014err}, the distortion of spatial modes also causes the decay of coherence and reduces the transmission rate of secure keys in free space quantum communication. In other words, the effect of turbulence on the propagation of OAM modes decides whether these modes can contribute to efficient quantum key generation over large distances.

Various aspects of OAM modes propagating through turbulence have been considered theoretically, including the detection probability of OAM modes \cite{Paterson2005, *Gopaul2007, *Tyler2009}, attenuation and crosstalk among multiple OAM channels \cite{Anguita2008}, the decay of entanglement for bipartite qubits \cite{Smith2006,Roux2011}, and the quantum channel capacity \cite{zhang2012}. A few experimental studies of the effect of turbulence on the OAM modes have also been reported \cite{pors,Ibrahim2013}. Some groups proposed methods to overcome the effect of turbulence on free space optical or quantum communication. These include, the use of post-processing (adaptive optics) \cite{leachao,*renetal} and pre-processing (robust states \cite{robust} and optimal encoding \cite{brun}) schemes. QKD protocols using qubit or multi-dimensional mutually unbiased bases (MUBs) encoded by means of OAM modes were recently tested experimentally \cite{Mafu2013,*vallone2014,*mir2014}. However, these studies only consider the effect of additional noise (such as turbulence) in the transmission channel, at a superficial level, if at all.

\begin{figure}[ht]
\includegraphics{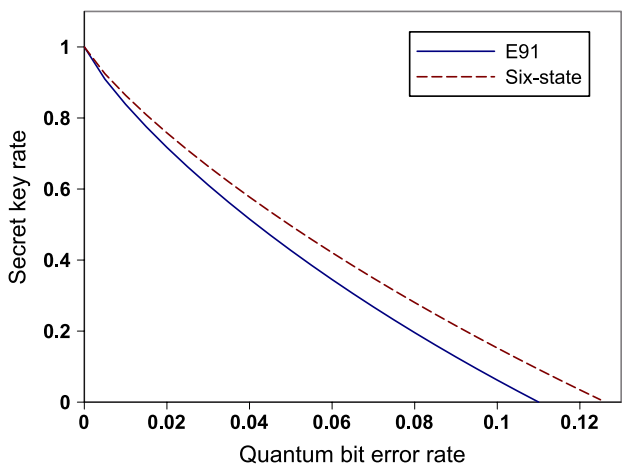}
\caption{(Color online) Comparison of the secret key rates of the E91 protocol and the six-state protocol, as a function of the quantum bit error rate.}
\label{rvsQ}
\end{figure}
 
In this article, we report on an experimental study of the influence of turbulence on the secret key rate for certain QKD protocols. For this purpose we prepare two photons in a maximally entangled state, using spontaneous parametric down-conversion (SPDC). We restrict the number of OAM modes to two (qubits) and simulate the effect of atmospheric turbulence on both photons through phase modulation by a single phase screen in each of the beam paths. Using the same setup as in \cite{Ibrahim2013}, we investigate qubits encoded with various combinations of OAM values. We study two QKD protocols: E91 \cite{Ekert1991} with different sets of MUBs and the six-state protocol \cite{Bruss1998}. These protocols are realized by performing measurements in the eigenbasis of the Pauli matrices on both photons. Since there are three MUBs for two dimensions one has three ways to select two sets of MUBs for the E91 protocol. For the six-state protocol we perform measurements in all three MUBs (the complete set for qubits) \cite{schwinger1960}. 

We find that secure QKD is possible over distances up to where entanglement decays to zero, but with lower secret key rate than entanglement of formation.

\section{Theory}
\label{theory}

For qubits (two-dimensional systems) the sets of eigenstates of the respective Pauli operators form MUBs. The three Pauli operators can be expressed as
\begin{align}
\hat{\sigma}_{\rm x} &= \ket{1}\bra{0}+\ket{0}\bra{1} \label{sigx} \\
\hat{\sigma}_{\rm y} &= i(\ket{1}\bra{0}-\ket{0}\bra{1}) \label{sigy} \\
\hat{\sigma}_{\rm z} &= \ket{0}\bra{0}-\ket{1}\bra{1} \label{sigz} .
\end{align}
The corresponding bases (sets of eigenstates) that are associated with the respective Pauli operators, are
\begin{align}
{\cal M}_x &= \left\{\frac{1}{\sqrt{2}}(\ket{0}\pm\ket{1})\right\} \label{mubx} \\
{\cal M}_y &= \left\{\frac{1}{\sqrt{2}}(\ket{0}\pm i\ket{1})\right\} \label{muby} \\
{\cal M}_z &= \left\{\ket{0},\ket{1}\right\} \label{mubz} .
\end{align}
These bases are mutually unbiased in that the measurement of a basis vector of one of these sets (say ${\cal M}_x$) in terms of another one of these bases (say ${\cal M}_y$) leads to equi-probable outcomes.

The first quantum key distribution protocol was developed by Bennett and Brassard in 1984 (BB84 protocol) \cite{Bennett1984}. The entanglement-based version of the BB84 protocol was developed by Ekert in 1991 (E91 protocol) \cite{Ekert1991}. In this protocol, the two parties, Alice and Bob, each obtain one subsystem of a maximally entangled state. In this study we use the Bell state
\begin{align}
\ket{\Phi^+} &= \frac{1}{\sqrt{2}}(\ket{01} + \ket{10}) .
\label{bell}
\end{align}
Each of the two parties, Alice and Bob, then randomly chooses a measurement basis from a set of two MUBs, $\{{\cal M}_1,{\cal M}_2\}$, which could be any two of the bases shown in Eqs.~(\ref{mubx}) to (\ref{mubz}), and perform a measurement on its respective subsystems in the chosen basis. The two parties repeat this process for a large number $n$ of quantum states and keep a record of their respective measurement outcomes, as well as the basis in which the measurements were performed. Alice and Bob, then publicly compare their measurement bases and keep only those results for which their bases matched, discarding the rest. This so-called sifting process would ideally result in an identical key for both parties.

Another QKD protocol of interest is the six-state protocol \cite{Bruss1998}, which is a straight-forward generalization of E91. The six-state protocol uses three orthonormal MUBs $\{{\cal M}_1,{\cal M}_2,{\cal M}_3\}$ [the three MUBs shown in Eqs.~(\ref{mubx}) to (\ref{mubz})], instead of just two. The rest of the protocol is analogous to E91. The six-state protocol produces a higher secret key rate than E91 for the same quantum bit error rate, as shown in Fig.~\ref{rvsQ}.

In the ideal scenario the protocols discussed above generate identical keys for Alice and Bob. However, differences between both keys can arise from disturbances due to eavesdropping, but also from imperfections in the state preparation, the transmission and the measurement process. It is rather difficult to differentiate between the error caused by eavesdropping and the errors of other origins. Therefore, it is safest to assign all the differences to eavesdropping. To estimate the average error, both parties compare a small portion of their measurements and calculate the so-called quantum bit error rate $Q$. It is given by the probability that Alice sends the state $\ket{\psi}$ and Bob projects, in an ideal measurement, onto an orthogonal state $\ket{\psi_\perp}$. In the entanglement-based protocols, the quantum bit error rate $Q$ is
\begin{align}
Q &= \frac{1}{\cal L} \sum_{\beta=1}^{\cal L} \sum_{k\ne k'} {\rm tr} \left( \ket{\psi^{\beta}_k}\bra{\psi^{\beta}_k} \otimes \ket{\psi^{\beta}_{k'}}\bra{\psi^{\beta}_{k'}} \rho_{_{AB}} \right)
\label{eqn-02}
\end{align}
where $\rho_{_{AB}}$ is the combined state of the system shared by Alice and Bob, whereas $\ket{\psi_k}$ and $\ket{\psi_{k'}}$ are the states on which their measurements project in case they chose the same basis. The index $\beta$ numerates the chosen bases and ${\cal L}$ is the number of bases available in the protocol:\ ${\cal L} = 2$ and ${\cal L} = 3$, for E91 and the six-state protocols, respectively.

The efficiency of the protocol is quantified by the secret key rate $r$, which represents the average number of secret key bits that can be distilled from each transmitted qubit by means of key sifting and privacy amplification \cite{Ferenczi2012}. The maximum number of secret key bits per transmitted two-level system is given by $r=1$. However, the expression of the minimum number of secret key bits $r_{\rm min}$ for a given value of the quantum bit error rate $Q$, for E91 (and BB84) reads \cite{Sheridan2010, Ferenczi2012}
\begin{align}
r_{\rm min} &= 1 + 2(1-Q)\log_2(1-Q) + 2Q\log_2\left(Q\right),\label{eqn-03}
\end{align}
and for the six-state protocol it is given by
\begin{align}
r_{\rm min} &= 1 + \frac{3}{2}Q\log_2\left(\frac{Q}{2}\right) + \left(1-\frac{3}{2}Q\right) \log_2\left(1-\frac{3}{2}Q\right).
\label{eqn-04}
\end{align}
In a comparison of key rates, the six-state protocol produces a higher secret key rate than E91 for the same quantum bit error rate, as shown in Fig. 2.

While the secret key rate is a measure for the efficiency of a particular QKD protocol, the amount of entanglement (quantified by the entanglement of formation) between the pairs of photons that arrive at Alice and Bob quantifies the quantum correlations that could be used to generate secret key bits shared by Alice and Bob \cite{Horodecki-review}. The entanglement of formation between two two-level photonic systems is calculated from the average joint density matrix of this bipartite system as \cite{Wootters1998},
\begin{align}
E = h\left(\frac{1+\sqrt{1-{\cal C}^2}}{2}\right),
\label{eof}
\end{align}
where $h(\cdot)$ is the binary entropy function, given by
\begin{align}
h(x) &= -x\log_2x - (1-x)\log_2(1-x) ,
\end{align}
and ${\cal C}$ is the concurrence, which is defined as
\begin{align}
{\cal C} &= \max \{0,\sqrt{\lambda_1}-\sqrt{\lambda_2}-\sqrt{\lambda_3}-\sqrt{\lambda_4}\} .
\label{conc}
\end{align}
In Eq.~(\ref{conc}) the $\lambda_n$'s are the eigenvalues, in decreasing order, of the matrix $\tilde\rho = \rho (\sigma_y\otimes\sigma_y) \rho^* (\sigma_y\otimes\sigma_y)$, where $^*$ represents the complex conjugate and $\sigma_y$ is the Pauli spin matrix.

\section{Experiment}
\label{experiment}

We implement these protocols in terms of the OAM degree of freedom of entangled photons. Entangled pairs of photons are prepared with the aid of SPDC, as explained below. These entangled pairs are then distributed through noisy channels (simulated turbulence) to two projective measurement setups (Alice and Bob). The projective measurements are made in terms of the helical modal basis (approximating Laguerre-Gaussian modes with radial index p=0). We'll denote this basis by $\{\ket{\ell}\}$ where $\ell$ represents the OAM index of the mode. For the purpose of the QKD protocols, we define the MUBs by replacing $\ket{0}\rightarrow\ket{{-\ell}}$ and $\ket{1}\rightarrow\ket{\ell}$ in Eqs.~(\ref{mubx}) to (\ref{mubz}), giving 
\begin{align}
{\cal M}_1 &= \left\{\ket{{-\ell}},\ket{\ell}\right\} \label{mub1} \\
{\cal M}_2 &= \left\{\frac{1}{\sqrt{2}}(\ket{{-\ell}}\pm\ket{\ell})\right\} \label{mub2} \\
{\cal M}_3 &= \left\{\frac{1}{\sqrt{2}}(\ket{{-\ell}}\pm i\ket{\ell})\right\} \label{mub3} .
\end{align}
We perform these measurements for $\ell$= 1, 3, 5 and 7.

Our aim to study the effect of atmospheric turbulence on the secret key rate $r$ is realized in the laboratory by simulating the propagation of the photons in a turbulent atmosphere, using spatial light modulators (SLMs). The influence of atmospheric turbulence on an optical beam, in weak scintillation conditions, can be simulated by an SLM that is encoded with a random phase function (as caused by fluctuations of the refractive index of turbulent air) in a single transversal plane of the beam \cite{Paterson2005}. To simulate the turbulent atmosphere, in agreement with the Kolmogorov theory of turbulence \cite{kolmogorov}, we compute the random phase function on the SLM as \cite{MF1, *MF2}
\begin{align}
\theta(x,y) &= \frac{k_0 \sqrt{2\pi L}}{\Delta_k} {\cal F}^{-1} \left\{ \chi(k_{\perp}) \sqrt{\Phi_n(|k_{\perp}|)} \right \} ,
\label{phscr}
\end{align}
where $k_0$ is the wavenumber ($=2\pi/\lambda$ where $\lambda$ is the wavelength of the light), $L$ is the propagation distance, $\Delta_k$ is the sampling interval in the frequency domain, ${\cal F}^{-1}\{\cdot\}$ is the two-dimensional inverse Fourier transform, $k_{\perp}$ is the two-dimensional wave vector in the transverse Fourier domain, and $\chi(k_{\perp})$ is a frequency domain delta-correlated zero-mean Gaussian pseudo-random complex function, obeying $\chi^*(k_{\perp}) = \chi(-k_{\perp})$, because $\theta(x, y)$ is real-valued. The refractive index power spectral density is expressed as \cite{kolmogorov, KNEPP1983, *Lane1992}
\begin{equation}
\Phi_n(k) = 0.033 C_n^2 k^{-11/3} ,
\label{eqn:kolmogorov}
\end{equation}
where $C_n^2$ is the refractive index structure constant, which determines the strength of the turbulence and $k$ is the magnitude of the spatial frequency vector.

\begin{figure}[ht]
\includegraphics{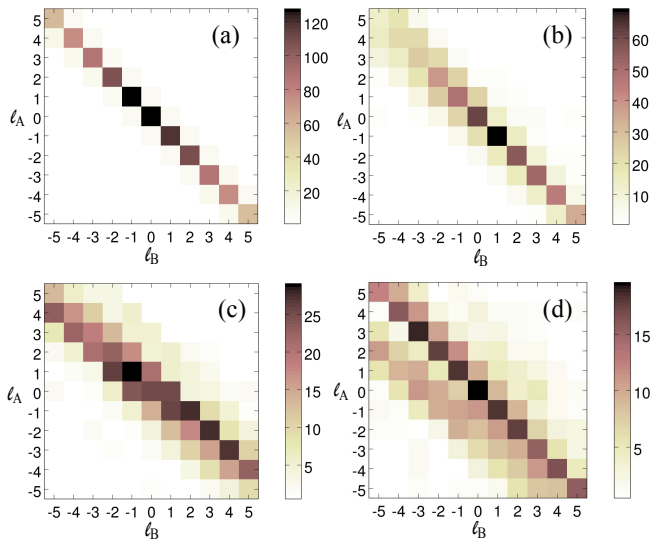}
\caption{(Color online) Mode scattering induced by turbulence. These figures show the coincidence counts as we measure simultaneously modes with $\ell_A$ in the signal beam and $\ell_B$ in the idler beam. (a) With no turbulence, only anti-correlated coincidences are observed. (b), (c) and (d) Progressively more severe turbulence conditions produce progressively more off-diagonal (uncorrelated) coincidence counts.}
\label{turboam}
\end{figure}

The distortion that is introduced by the modulation of the optical beams by the random phase functions on the SLMs leads to crosstalk between OAM modes. One can observe this crosstalk in terms of the decay of correlations between the OAM modes for a down-converted pair of photons. This is demonstrated by the graphs in Fig.~\ref{turboam}, which show how the anti-correlation between the OAM index of the two down-converted photons deteriorates with increasing scintillation, as introduced by the random phase function on the SLMs.

In our simulated turbulence experiment, the turbulence strength is combined with the wavelength and the propagation distance into the Fried parameter, which, for plane waves, is given by
\begin{equation}
r_0 = 0.185\left(\frac{\lambda^2}{C_n^2 L} \right)^{3/5} .
\label{defr0}
\end{equation}
One can define a dimensionless quantity
\begin{equation}
{\cal W}=\frac{w_0}{r_0} ,
\label{defW}
\end{equation}
where $w_0$ is the beam radius. Since ${\cal W}$ contains the propagation distance, it can be regarded as an indication of scintillation strength rather than turbulence strength. It was found \cite{Smith2006} that, under weak scintillation conditions, the evolution of the entanglement of a photon pair that is entangled in its spatial degrees of freedom (such as OAM), is completely determined by ${\cal W}$. In our experiment we assume weak scintillation and simulate the turbulence with a single phase screen. As a result the parameter that we use to quantify the strength of the scintillation/turbulence is ${\cal W}$.

\begin{figure}[ht]
\centerline{\includegraphics{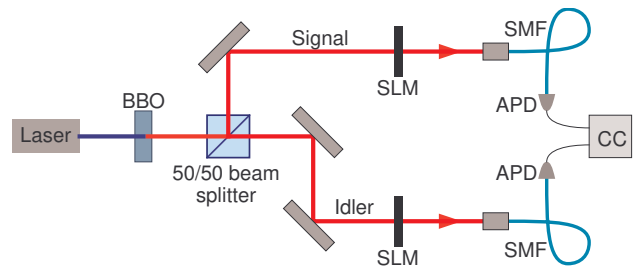}}
\caption{(Color online) Experimental setup to generate and measure entangled photon pairs. A UV laser source pumps a type-I BBO crystal to produce pairs of entangled photons via SPDC. The crystal plane is imaged onto the SLMs and each SLM plane is imaged onto the input of a SMF.}
\label{setup}
\end{figure}

\begin{figure*}[ht]
\centering
\includegraphics{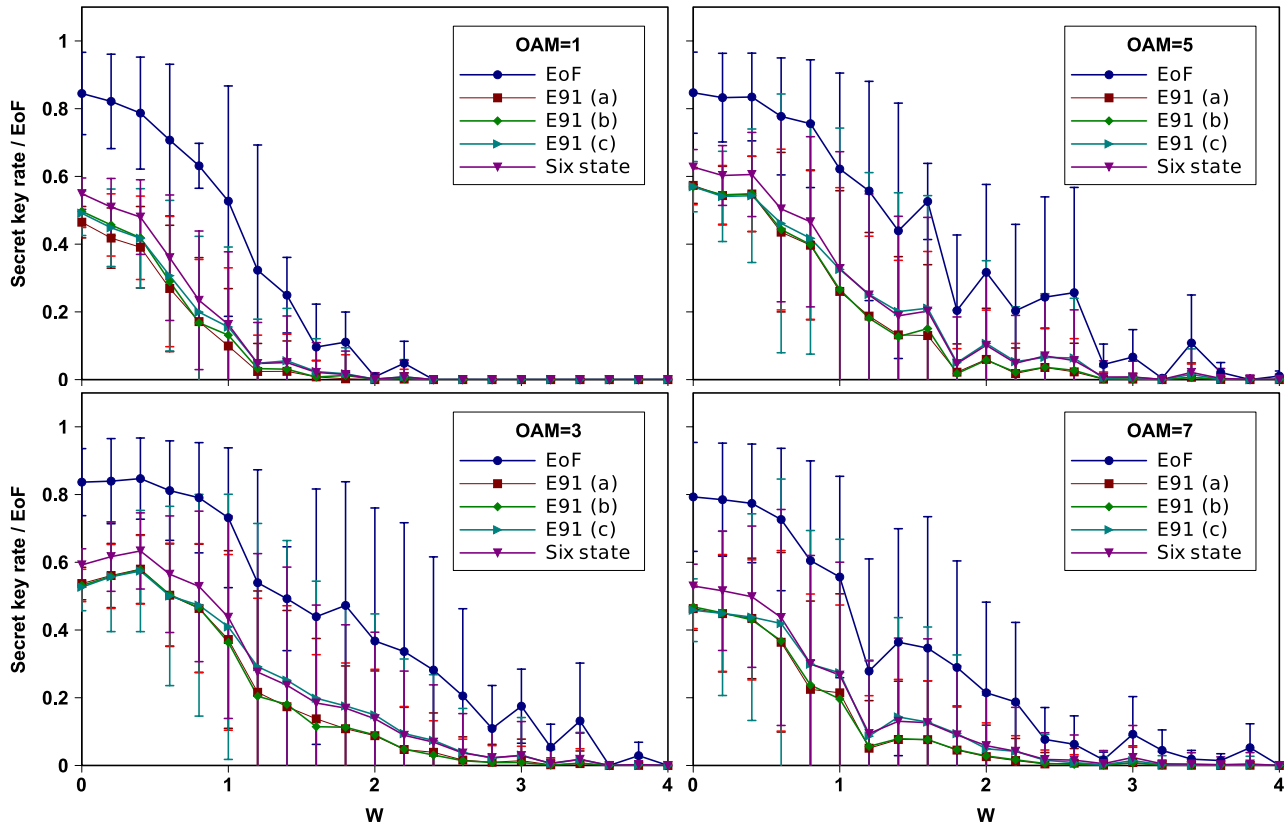}
\caption{(Color online) Measured secret key rates of E91 for three different possible choices of the two bases and six-state protocol as a function of ${\cal W}=w_0/r_0$ for different OAM encodings.}
\label{E91vs6state}
\end{figure*}

Our experimental setup is shown diagrammatically in Fig.~\ref{setup}. It is identical to the one used in \cite{Ibrahim2013}. Entangled photon pairs are generated, using SPDC, by pumping a type-I Barium borate (BBO) crystal with a 355~nm laser. The collinear, degenerate down-converted photons are imaged via a 4f-system from the BBO crystal plane to the two separate SLMs. These HoloEye SLMs consist of $1920\times 1080$ pixels, with a pixel size of 8~$\mu$m. The photon pairs are projected onto particular helical modes, depending on the helical phase functions that are encoded onto the SLMs. The expression for the transmission (reflection) function of the SLMs encoded with these helical phase functions is
\begin{equation}
t(\phi)=\exp(i\ell\phi) ,
\end{equation}
where $\phi$ is the azimuthal coordinate and $\ell$ is the OAM index (an integer). Random phase functions are added to each SLM to simulate propagation through atmospheric turbulence. The resulting optical fields after the SLMs are then imaged via a 4f-system onto the input facet of single-mode fibers (SMFs). The single photons are detected with avalanche photo diodes (APDs), which are connected to a coincidence counter (CC) with a gating time of 12.5~ns.

\section{Results and discussion}
\label{result} 

Using the projective measurements in the helical basis, we perform quantum state tomography to reproduce the density matrices for the observed quantum states in the two-qubit Hilbert spaces given by the particular value of $\ell$. Such a quantum state tomography is performed by cycling through the different bases composed of the OAM modes in the restricted Hilbert space on each SLM and recording the coincidence count rates. This quantum state tomography is repeated 30 times, with different sets of random phase functions (different realizations of the turbulent medium), for each value of ${\cal W}$, within the range $0\leq{\cal W}\leq 4$, and for each of the four values of $\ell$=1, 3, 5, and 7. The resulting density matrices are averaged to obtain a mean density matrix for a given ${\cal W}$ and $\ell$.

These mean density matrices are used to calculate the quantum bit error rate $Q$, with the aid of Eq.~\eqref{eqn-02}, and then the minimal secret key rates $r_{\rm min}$ for the E91 protocol, using Eq.~\eqref{eqn-03}, and for the six-state protocol, using Eq.~\eqref{eqn-04}. The mean density matrices are also used to compute the entanglement of formation with the aid of Eq.~\eqref{eof}. These minimal secret key rates $r_{\rm min}$ are shown, together with the entanglement of formation (EoF), as a function of ${\cal W}$ in Fig.~\ref{E91vs6state}. The calculation of $r_{\rm min}$ for the E91 protocol is done for all three different ways of choosing the two MUBs from those in Eqs.~(\ref{mub1}) to (\ref{mub3}). The curves denoted by E91 (a), (b) and (c) in Fig.~\ref{E91vs6state} represent the three cases where the chosen MUBs are $\{{\cal M}_1,{\cal M}_3\}$, $\{{\cal M}_1,{\cal M}_2\}$ and $\{{\cal M}_2,{\cal M}_3\}$, respectively. All the MUBs in Eqs.~(\ref{mub1}) to (\ref{mub3}) are used for the six-state protocol.

Although the qualitative behavior of $r_{\rm min}$ and the EoF is the same, the former is lower than the latter and the former also decays to zero at a smaller value of ${\cal W}$ than the latter. This could be caused by a non-optimal choice of the measurement bases in the experiment. By choosing bases that are more optimal, one may be able to increase the minimal secret key rate $r_{\rm min}$.

Comparing the secret key rate of the different QKD protocols, as shown in Fig.~\ref{E91vs6state}, we find that the secret key rate in the E91 protocols are in general lower than that of the six-state protocol. This result is to be expected, since the six-state protocol is known to produce a higher secret key rate than E91 for the same quantum bit error rate (see Fig.~\ref{rvsQ}). Moreover, the choice of MUBs for the E91 protocol makes a slight difference in the performance:\ the set $\{{\cal M}_2,{\cal M}_3\}$ gives slightly better performance than the other two choices for larger values of ${\cal W}$.

The aim of our experiment is to simulate the effect of atmospheric turbulence on down-converted pairs of photons, which are entangled in their spatial degrees of freedom. This allows us to study the security of QKD protocols that employ this type of entanglement resource in free space communication. Under weak scintillation conditions (as assumed in our experiment), the distance scale at which the concurrence decays to zero can be estimated from the observation (see Fig.~\ref{E91vs6state}) that, up to some numerical factor of order 1, the entanglement of formation decays to zero when ${\cal W}\approx 1$. According to numerical simulation results \cite{Ibrahim2013}, the relationship is actually ${\cal W}\approx \sqrt{\ell}$. The resulting concurrence decay distance scale is then given by \cite{Ibrahim2013}
\begin{align}
L_{\rm dec} \approx \frac{0.06\lambda^2\ell^{5/6}}{w_0^{5/3}C_n^2} ,
\label{decay}
\end{align}
where the numerical constant can vary within an order of magnitude. Hence, the distance over which QKD can be operated successfully through a free space channel, depends on the dimension parameters and the OAM index, as shown in Eq.~\eqref{decay}. 

As an example, consider the challenge of replicating a previous reported free space QKD experiment with polarization entangled photons over a distance of 144 km \cite{Ursin2007etal}. Here $C^{2}_{n} \approx 5 \times 10^{-16}$~m$^{-2/3}$ and $w_0 \approx 50$~mm for a photon wavelength of 710~nm. If an $\ell = 1$ qubit basis was used, the transmission distance for the same experimental parameters would be a little less than 10 km, or roughly an order of magnitude less than for polarization. The 144 km distance could only be reached (under the same experimental conditions) by using large OAM values, $\ell > 25$. This reinforces the very motivation for OAM as a basis for encoding information: its value lies in realizing higher dimensional states and not in replicating qubit states. 

It should be noted that distance estimations are here made under the assumption of weak scintillation conditions, which require that the distance is smaller than the Rayleigh range. However, by varying some of the parameters, one can obtain a distance that is larger than the Rayleigh range, in which case it would extend into a region where the weak scintillation condition does not apply anymore. As a result the predicted distance would not be reliable. To obtain reliable predictions of operating distances under strong scintillation conditions, one needs to employ multiple phase screen methods \cite{Roux2011}.

\section{Summary}
\label{summary}

We performed an experimental study of the effect of atmospheric turbulence on the security of certain QKD protocols: E91 with different choices of MUBs and the six-state protocol. We found that, in the weak scintillation limit, one can distribute a quantum key securely over distances comparable to those over which the entanglement survives, but at a slightly lower secret key rate compared to the entanglement of formation. 


\begin{thebibliography}{53}%
\makeatletter
\providecommand \@ifxundefined [1]{%
 \@ifx{#1\undefined}
}%
\providecommand \@ifnum [1]{%
 \ifnum #1\expandafter \@firstoftwo
 \else \expandafter \@secondoftwo
 \fi
}%
\providecommand \@ifx [1]{%
 \ifx #1\expandafter \@firstoftwo
 \else \expandafter \@secondoftwo
 \fi
}%
\providecommand \natexlab [1]{#1}%
\providecommand \enquote  [1]{``#1''}%
\providecommand \bibnamefont  [1]{#1}%
\providecommand \bibfnamefont [1]{#1}%
\providecommand \citenamefont [1]{#1}%
\providecommand \href@noop [0]{\@secondoftwo}%
\providecommand \href [0]{\begingroup \@sanitize@url \@href}%
\providecommand \@href[1]{\@@startlink{#1}\@@href}%
\providecommand \@@href[1]{\endgroup#1\@@endlink}%
\providecommand \@sanitize@url [0]{\catcode `\\12\catcode `\$12\catcode
  `\&12\catcode `\#12\catcode `\^12\catcode `\_12\catcode `\%12\relax}%
\providecommand \@@startlink[1]{}%
\providecommand \@@endlink[0]{}%
\providecommand \url  [0]{\begingroup\@sanitize@url \@url }%
\providecommand \@url [1]{\endgroup\@href {#1}{\urlprefix }}%
\providecommand \urlprefix  [0]{URL }%
\providecommand \Eprint [0]{\href }%
\providecommand \doibase [0]{http://dx.doi.org/}%
\providecommand \selectlanguage [0]{\@gobble}%
\providecommand \bibinfo  [0]{\@secondoftwo}%
\providecommand \bibfield  [0]{\@secondoftwo}%
\providecommand \translation [1]{[#1]}%
\providecommand \BibitemOpen [0]{}%
\providecommand \bibitemStop [0]{}%
\providecommand \bibitemNoStop [0]{.\EOS\space}%
\providecommand \EOS [0]{\spacefactor3000\relax}%
\providecommand \BibitemShut  [1]{\csname bibitem#1\endcsname}%
\let\auto@bib@innerbib\@empty
\bibitem [{\citenamefont {Lo}(1998)}]{Lo1998}%
  \BibitemOpen
  \bibfield  {author} {\bibinfo {author} {\bibfnamefont {H.-K.}\ \bibnamefont
  {Lo}},\ }\href@noop {} {\emph {\bibinfo {title} {Introduction to Quantum
  Computation and Information}}},\ edited by\ \bibinfo {editor} {\bibfnamefont
  {H.-K.}\ \bibnamefont {Lo}}, \bibinfo {editor} {\bibfnamefont
  {S.}~\bibnamefont {Popescu}}, \ and\ \bibinfo {editor} {\bibfnamefont
  {T.}~\bibnamefont {Spiller}}\ (\bibinfo  {publisher} {World Scientific,
  Singapore},\ \bibinfo {year} {1998})\BibitemShut {NoStop}%
\bibitem [{\citenamefont {Ekert}\ \emph {et~al.}(2001)\citenamefont {Ekert},
  \citenamefont {Gisin}, \citenamefont {Huttner}, \citenamefont {Inamori},\
  and\ \citenamefont {Weinfurter}}]{Ekert2001}%
  \BibitemOpen
  \bibfield  {author} {\bibinfo {author} {\bibfnamefont {A.~K.}\ \bibnamefont
  {Ekert}}, \bibinfo {author} {\bibfnamefont {N.}~\bibnamefont {Gisin}},
  \bibinfo {author} {\bibfnamefont {B.}~\bibnamefont {Huttner}}, \bibinfo
  {author} {\bibfnamefont {H.}~\bibnamefont {Inamori}}, \ and\ \bibinfo
  {author} {\bibfnamefont {H.}~\bibnamefont {Weinfurter}},\ }\href@noop {}
  {\emph {\bibinfo {title} {The Physics of Quantum information}}},\ edited by\
  \bibinfo {editor} {\bibfnamefont {D.}~\bibnamefont {Bouwmeester}}, \bibinfo
  {editor} {\bibfnamefont {A.~K.}\ \bibnamefont {Ekert}}, \ and\ \bibinfo
  {editor} {\bibfnamefont {A.}~\bibnamefont {Zeilinger}}\ (\bibinfo
  {publisher} {Springer Verlag, London},\ \bibinfo {year} {2001})\BibitemShut
  {NoStop}%
\bibitem [{\citenamefont {Gisin}\ \emph {et~al.}(2002)\citenamefont {Gisin},
  \citenamefont {Ribordy}, \citenamefont {Tittel},\ and\ \citenamefont
  {Zbinden}}]{Gisin2002}%
  \BibitemOpen
  \bibfield  {author} {\bibinfo {author} {\bibfnamefont {N.}~\bibnamefont
  {Gisin}}, \bibinfo {author} {\bibfnamefont {G.}~\bibnamefont {Ribordy}},
  \bibinfo {author} {\bibfnamefont {W.}~\bibnamefont {Tittel}}, \ and\ \bibinfo
  {author} {\bibfnamefont {H.}~\bibnamefont {Zbinden}},\ }\href@noop {}
  {\bibfield  {journal} {\bibinfo  {journal} {Rev. Mod. Phys.}\ }\textbf
  {\bibinfo {volume} {74}},\ \bibinfo {pages} {145} (\bibinfo {year}
  {2002})}\BibitemShut {NoStop}%
\bibitem [{\citenamefont {Scarani}(2006)}]{Scarani2006}%
  \BibitemOpen
  \bibfield  {author} {\bibinfo {author} {\bibfnamefont {V.}~\bibnamefont
  {Scarani}},\ }\href@noop {} {\emph {\bibinfo {title} {Quantum Physics-- A
  First Encounter}}}\ (\bibinfo  {publisher} {Oxford University Press,
  Oxford},\ \bibinfo {year} {2006})\BibitemShut {NoStop}%
\bibitem [{\citenamefont {Bellac}(2006)}]{Le2006}%
  \BibitemOpen
  \bibfield  {author} {\bibinfo {author} {\bibfnamefont {M.~L.}\ \bibnamefont
  {Bellac}},\ }\href@noop {} {\emph {\bibinfo {title} {A Short Introduction to
  Quantum Information and Quantum Computation}}}\ (\bibinfo  {publisher}
  {Cambridge University Press, Cambridge},\ \bibinfo {year} {2006})\BibitemShut
  {NoStop}%
\bibitem [{\citenamefont {Du$\breve{s}$ek}\ \emph {et~al.}(2006)\citenamefont
  {Du$\breve{s}$ek}, \citenamefont {L\"utkenhaus},\ and\ \citenamefont
  {Hendrych}}]{Dusek2006}%
  \BibitemOpen
  \bibfield  {author} {\bibinfo {author} {\bibfnamefont {M.}~\bibnamefont
  {Du$\breve{s}$ek}}, \bibinfo {author} {\bibfnamefont {N.}~\bibnamefont
  {L\"utkenhaus}}, \ and\ \bibinfo {author} {\bibfnamefont {M.}~\bibnamefont
  {Hendrych}},\ }\href@noop {} {\bibfield  {journal} {\bibinfo  {journal}
  {Progress in Optics}\ }\textbf {\bibinfo {volume} {49}},\ \bibinfo {pages}
  {381} (\bibinfo {year} {2006})}\BibitemShut {NoStop}%
\bibitem [{\citenamefont {Lo}\ and\ \citenamefont {Zhao}(2009)}]{Lo2009}%
  \BibitemOpen
  \bibfield  {author} {\bibinfo {author} {\bibfnamefont {H.-K.}\ \bibnamefont
  {Lo}}\ and\ \bibinfo {author} {\bibfnamefont {Y.}~\bibnamefont {Zhao}},\
  }\href@noop {} {\bibfield  {journal} {\bibinfo  {journal} {Encyclopedia of
  Complexity and System Science}\ }\textbf {\bibinfo {volume} {8}},\ \bibinfo
  {pages} {7265} (\bibinfo {year} {2009})}\BibitemShut {NoStop}%
\bibitem [{\citenamefont {Scarani}\ \emph {et~al.}(2009)\citenamefont
  {Scarani}, \citenamefont {Bechmann-Pasquinucci}, \citenamefont {Cerf},
  \citenamefont {Du\ifmmode~\check{s}\else \v{s}\fi{}ek}, \citenamefont
  {L\"{u}tkenhaus},\ and\ \citenamefont {Peev}}]{Scarani2009}%
  \BibitemOpen
  \bibfield  {author} {\bibinfo {author} {\bibfnamefont {V.}~\bibnamefont
  {Scarani}}, \bibinfo {author} {\bibfnamefont {H.}~\bibnamefont
  {Bechmann-Pasquinucci}}, \bibinfo {author} {\bibfnamefont {N.~J.}\
  \bibnamefont {Cerf}}, \bibinfo {author} {\bibfnamefont {M.}~\bibnamefont
  {Du\ifmmode~\check{s}\else \v{s}\fi{}ek}}, \bibinfo {author} {\bibfnamefont
  {N.}~\bibnamefont {L\"{u}tkenhaus}}, \ and\ \bibinfo {author} {\bibfnamefont
  {M.}~\bibnamefont {Peev}},\ }\href@noop {} {\bibfield  {journal} {\bibinfo
  {journal} {Rev. Mod. Phys.}\ }\textbf {\bibinfo {volume} {81}},\ \bibinfo
  {pages} {1301} (\bibinfo {year} {2009})}\BibitemShut {NoStop}%
\bibitem [{\citenamefont {Gr{\"o}blacher}\ \emph {et~al.}(2006)\citenamefont
  {Gr{\"o}blacher}, \citenamefont {Jennewein}, \citenamefont {Vaziri},
  \citenamefont {Weihs},\ and\ \citenamefont {Zeilinger}}]{groblacher2006}%
  \BibitemOpen
  \bibfield  {author} {\bibinfo {author} {\bibfnamefont {S.}~\bibnamefont
  {Gr{\"o}blacher}}, \bibinfo {author} {\bibfnamefont {T.}~\bibnamefont
  {Jennewein}}, \bibinfo {author} {\bibfnamefont {A.}~\bibnamefont {Vaziri}},
  \bibinfo {author} {\bibfnamefont {G.}~\bibnamefont {Weihs}}, \ and\ \bibinfo
  {author} {\bibfnamefont {A.}~\bibnamefont {Zeilinger}},\ }\href@noop {}
  {\bibfield  {journal} {\bibinfo  {journal} {New Journal of Physics}\ }\textbf
  {\bibinfo {volume} {8}},\ \bibinfo {pages} {75} (\bibinfo {year}
  {2006})}\BibitemShut {NoStop}%
\bibitem [{\citenamefont {Ursin}\ \emph {et~al.}(2007)\citenamefont {Ursin},
  \citenamefont {Tiefenbacher}, \citenamefont {Schmitt-Manderbach},
  \citenamefont {Weier} \emph {et~al.}}]{Ursin2007etal}%
  \BibitemOpen
  \bibfield  {author} {\bibinfo {author} {\bibfnamefont {R.}~\bibnamefont
  {Ursin}}, \bibinfo {author} {\bibfnamefont {F.}~\bibnamefont {Tiefenbacher}},
  \bibinfo {author} {\bibfnamefont {T.}~\bibnamefont {Schmitt-Manderbach}},
  \bibinfo {author} {\bibfnamefont {H.}~\bibnamefont {Weier}},  \emph
  {et~al.},\ }\href@noop {} {\bibfield  {journal} {\bibinfo  {journal} {Nature
  Phys.}\ }\textbf {\bibinfo {volume} {3}},\ \bibinfo {pages} {481} (\bibinfo
  {year} {2007})}\BibitemShut {NoStop}%
\bibitem [{\citenamefont {Erven}\ \emph {et~al.}(2008)\citenamefont {Erven},
  \citenamefont {Couteau}, \citenamefont {Laflamme},\ and\ \citenamefont
  {Weihs}}]{Erven2008}%
  \BibitemOpen
  \bibfield  {author} {\bibinfo {author} {\bibfnamefont {C.}~\bibnamefont
  {Erven}}, \bibinfo {author} {\bibfnamefont {C.}~\bibnamefont {Couteau}},
  \bibinfo {author} {\bibfnamefont {R.}~\bibnamefont {Laflamme}}, \ and\
  \bibinfo {author} {\bibfnamefont {G.}~\bibnamefont {Weihs}},\ }\href@noop {}
  {\bibfield  {journal} {\bibinfo  {journal} {Opt. Express}\ }\textbf {\bibinfo
  {volume} {16}},\ \bibinfo {pages} {16840} (\bibinfo {year}
  {2008})}\BibitemShut {NoStop}%
\bibitem [{\citenamefont {Peev}\ \emph {et~al.}(2009)\citenamefont {Peev},
  \citenamefont {Pacher}, \citenamefont {All\'eaume}, \citenamefont {Barreiro}
  \emph {et~al.}}]{Peev2009etal}%
  \BibitemOpen
  \bibfield  {author} {\bibinfo {author} {\bibfnamefont {M.}~\bibnamefont
  {Peev}}, \bibinfo {author} {\bibfnamefont {C.}~\bibnamefont {Pacher}},
  \bibinfo {author} {\bibfnamefont {R.}~\bibnamefont {All\'eaume}}, \bibinfo
  {author} {\bibfnamefont {C.}~\bibnamefont {Barreiro}},  \emph {et~al.},\
  }\href@noop {} {\bibfield  {journal} {\bibinfo  {journal} {New J. Phys.}\
  }\textbf {\bibinfo {volume} {11}},\ \bibinfo {pages} {075001} (\bibinfo
  {year} {2009})}\BibitemShut {NoStop}%
\bibitem [{\citenamefont {Peloso}\ \emph {et~al.}(2009)\citenamefont {Peloso},
  \citenamefont {Gerhardt}, \citenamefont {Ho}, \citenamefont {Lamas-Linares},\
  and\ \citenamefont {Kurtsiefer}}]{Peloso2009}%
  \BibitemOpen
  \bibfield  {author} {\bibinfo {author} {\bibfnamefont {M.~P.}\ \bibnamefont
  {Peloso}}, \bibinfo {author} {\bibfnamefont {I.}~\bibnamefont {Gerhardt}},
  \bibinfo {author} {\bibfnamefont {C.}~\bibnamefont {Ho}}, \bibinfo {author}
  {\bibfnamefont {A.}~\bibnamefont {Lamas-Linares}}, \ and\ \bibinfo {author}
  {\bibfnamefont {C.}~\bibnamefont {Kurtsiefer}},\ }\href@noop {} {\bibfield
  {journal} {\bibinfo  {journal} {New J. Phys.}\ }\textbf {\bibinfo {volume}
  {11}},\ \bibinfo {pages} {045007} (\bibinfo {year} {2009})}\BibitemShut
  {NoStop}%
\bibitem [{\citenamefont {Jin}\ \emph {et~al.}(2010)\citenamefont {Jin},
  \citenamefont {Ren}, \citenamefont {Yang}, \citenamefont {Yi}, \citenamefont
  {Zhou}, \citenamefont {Xu}, \citenamefont {Wang}, \citenamefont {Yang},
  \citenamefont {Hu}, \citenamefont {Jiang}, \citenamefont {Yang},
  \citenamefont {Yin}, \citenamefont {Chen}, \citenamefont {Peng},\ and\
  \citenamefont {Pan}}]{Jin2010}%
  \BibitemOpen
  \bibfield  {author} {\bibinfo {author} {\bibfnamefont {X.-M.}\ \bibnamefont
  {Jin}}, \bibinfo {author} {\bibfnamefont {J.-G.}\ \bibnamefont {Ren}},
  \bibinfo {author} {\bibfnamefont {B.}~\bibnamefont {Yang}}, \bibinfo {author}
  {\bibfnamefont {Z.-H.}\ \bibnamefont {Yi}}, \bibinfo {author} {\bibfnamefont
  {F.}~\bibnamefont {Zhou}}, \bibinfo {author} {\bibfnamefont {X.-F.}\
  \bibnamefont {Xu}}, \bibinfo {author} {\bibfnamefont {S.-K.}\ \bibnamefont
  {Wang}}, \bibinfo {author} {\bibfnamefont {D.}~\bibnamefont {Yang}}, \bibinfo
  {author} {\bibfnamefont {Y.-F.}\ \bibnamefont {Hu}}, \bibinfo {author}
  {\bibfnamefont {S.}~\bibnamefont {Jiang}}, \bibinfo {author} {\bibfnamefont
  {T.}~\bibnamefont {Yang}}, \bibinfo {author} {\bibfnamefont {H.}~\bibnamefont
  {Yin}}, \bibinfo {author} {\bibfnamefont {K.}~\bibnamefont {Chen}}, \bibinfo
  {author} {\bibfnamefont {C.-Z.}\ \bibnamefont {Peng}}, \ and\ \bibinfo
  {author} {\bibfnamefont {J.-W.}\ \bibnamefont {Pan}},\ }\href@noop {}
  {\bibfield  {journal} {\bibinfo  {journal} {Nat. Photon.}\ }\textbf {\bibinfo
  {volume} {4}},\ \bibinfo {pages} {376} (\bibinfo {year} {2010})}\BibitemShut
  {NoStop}%
\bibitem [{\citenamefont {Sasaki}\ \emph {et~al.}(2011)\citenamefont {Sasaki},
  \citenamefont {Fujiwara}, \citenamefont {Ishizuka}, \citenamefont {Klaus}
  \emph {et~al.}}]{Sasaki2011etal}%
  \BibitemOpen
  \bibfield  {author} {\bibinfo {author} {\bibfnamefont {M.}~\bibnamefont
  {Sasaki}}, \bibinfo {author} {\bibfnamefont {M.}~\bibnamefont {Fujiwara}},
  \bibinfo {author} {\bibfnamefont {H.}~\bibnamefont {Ishizuka}}, \bibinfo
  {author} {\bibfnamefont {W.}~\bibnamefont {Klaus}},  \emph {et~al.},\
  }\href@noop {} {\bibfield  {journal} {\bibinfo  {journal} {Opt. Express}\
  }\textbf {\bibinfo {volume} {19}},\ \bibinfo {pages} {10387} (\bibinfo {year}
  {2011})}\BibitemShut {NoStop}%
\bibitem [{\citenamefont {Erven}\ \emph {et~al.}(2012)\citenamefont {Erven},
  \citenamefont {Heim}, \citenamefont {Meyer-Scott}, \citenamefont {Bourgoin},
  \citenamefont {Laflamme}, \citenamefont {Weihs},\ and\ \citenamefont
  {Jennewein}}]{Erven2012}%
  \BibitemOpen
  \bibfield  {author} {\bibinfo {author} {\bibfnamefont {C.}~\bibnamefont
  {Erven}}, \bibinfo {author} {\bibfnamefont {B.}~\bibnamefont {Heim}},
  \bibinfo {author} {\bibfnamefont {E.}~\bibnamefont {Meyer-Scott}}, \bibinfo
  {author} {\bibfnamefont {J.~P.}\ \bibnamefont {Bourgoin}}, \bibinfo {author}
  {\bibfnamefont {R.}~\bibnamefont {Laflamme}}, \bibinfo {author}
  {\bibfnamefont {G.}~\bibnamefont {Weihs}}, \ and\ \bibinfo {author}
  {\bibfnamefont {T.}~\bibnamefont {Jennewein}},\ }\href@noop {} {\bibfield
  {journal} {\bibinfo  {journal} {New J. Phys.}\ }\textbf {\bibinfo {volume}
  {14}},\ \bibinfo {pages} {123018} (\bibinfo {year} {2012})}\BibitemShut
  {NoStop}%
\bibitem [{Note1()}]{Note1}%
  \BibitemOpen
  \bibinfo {note} {Quintessence~Labs~(www.quintessencelabs.com), MagiQ
  Technologies~(www.magiqtech.com), idQuantique~(www.idquantique.com), and
  SeQureNet~(www.sequrenet.fr)}\BibitemShut {NoStop}%
\bibitem [{\citenamefont {Dixon}\ \emph {et~al.}(2008)\citenamefont {Dixon},
  \citenamefont {Yuan}, \citenamefont {Dynes}, \citenamefont {Sharpe},\ and\
  \citenamefont {Shields}}]{Dixon2008}%
  \BibitemOpen
  \bibfield  {author} {\bibinfo {author} {\bibfnamefont {A.~R.}\ \bibnamefont
  {Dixon}}, \bibinfo {author} {\bibfnamefont {Z.~L.}\ \bibnamefont {Yuan}},
  \bibinfo {author} {\bibfnamefont {J.~F.}\ \bibnamefont {Dynes}}, \bibinfo
  {author} {\bibfnamefont {A.~W.}\ \bibnamefont {Sharpe}}, \ and\ \bibinfo
  {author} {\bibfnamefont {A.~J.}\ \bibnamefont {Shields}},\ }\href@noop {}
  {\bibfield  {journal} {\bibinfo  {journal} {Opt. Express}\ }\textbf {\bibinfo
  {volume} {16}},\ \bibinfo {pages} {18790} (\bibinfo {year}
  {2008})}\BibitemShut {NoStop}%
\bibitem [{\citenamefont {Boone}\ \emph {et~al.}(2014)\citenamefont {Boone},
  \citenamefont {Bourgoin}, \citenamefont {Meyer-Scott}, \citenamefont
  {Heshami}, \citenamefont {Jennewein},\ and\ \citenamefont
  {Simon}}]{boone2014}%
  \BibitemOpen
  \bibfield  {author} {\bibinfo {author} {\bibfnamefont {K.}~\bibnamefont
  {Boone}}, \bibinfo {author} {\bibfnamefont {J.-P.}\ \bibnamefont {Bourgoin}},
  \bibinfo {author} {\bibfnamefont {E.}~\bibnamefont {Meyer-Scott}}, \bibinfo
  {author} {\bibfnamefont {K.}~\bibnamefont {Heshami}}, \bibinfo {author}
  {\bibfnamefont {T.}~\bibnamefont {Jennewein}}, \ and\ \bibinfo {author}
  {\bibfnamefont {C.}~\bibnamefont {Simon}},\ }\href@noop {} {\bibfield
  {journal} {\bibinfo  {journal} {arXiv preprint arXiv:1410.5384}\ } (\bibinfo
  {year} {2014})}\BibitemShut {NoStop}%
\bibitem [{\citenamefont {Allen}\ \emph {et~al.}(1992)\citenamefont {Allen},
  \citenamefont {Beijersbergen}, \citenamefont {Spreeuw},\ and\ \citenamefont
  {Woerdman}}]{Allen1992}%
  \BibitemOpen
  \bibfield  {author} {\bibinfo {author} {\bibfnamefont {L.}~\bibnamefont
  {Allen}}, \bibinfo {author} {\bibfnamefont {M.}~\bibnamefont
  {Beijersbergen}}, \bibinfo {author} {\bibfnamefont {R.}~\bibnamefont
  {Spreeuw}}, \ and\ \bibinfo {author} {\bibfnamefont {J.}~\bibnamefont
  {Woerdman}},\ }\href@noop {} {\bibfield  {journal} {\bibinfo  {journal}
  {\pra}\ }\textbf {\bibinfo {volume} {45}},\ \bibinfo {pages} {8185} (\bibinfo
  {year} {1992})}\BibitemShut {NoStop}%
\bibitem [{\citenamefont {Malik}\ \emph {et~al.}(2012)\citenamefont {Malik},
  \citenamefont {O'Sullivan}, \citenamefont {Rodenburg}, \citenamefont
  {Mirhosseini}, \citenamefont {Leach}, \citenamefont {Lavery}, \citenamefont
  {Padgett},\ and\ \citenamefont {Boyd}}]{malik2012}%
  \BibitemOpen
  \bibfield  {author} {\bibinfo {author} {\bibfnamefont {M.}~\bibnamefont
  {Malik}}, \bibinfo {author} {\bibfnamefont {M.}~\bibnamefont {O'Sullivan}},
  \bibinfo {author} {\bibfnamefont {B.}~\bibnamefont {Rodenburg}}, \bibinfo
  {author} {\bibfnamefont {M.}~\bibnamefont {Mirhosseini}}, \bibinfo {author}
  {\bibfnamefont {J.}~\bibnamefont {Leach}}, \bibinfo {author} {\bibfnamefont
  {M.~P.}\ \bibnamefont {Lavery}}, \bibinfo {author} {\bibfnamefont {M.~J.}\
  \bibnamefont {Padgett}}, \ and\ \bibinfo {author} {\bibfnamefont {R.~W.}\
  \bibnamefont {Boyd}},\ }\href@noop {} {\bibfield  {journal} {\bibinfo
  {journal} {Opt. Express}\ }\textbf {\bibinfo {volume} {20}},\ \bibinfo
  {pages} {13195} (\bibinfo {year} {2012})}\BibitemShut {NoStop}%
\bibitem [{\citenamefont {Rodenburg}\ \emph {et~al.}(2012)\citenamefont
  {Rodenburg}, \citenamefont {Lavery}, \citenamefont {Malik}, \citenamefont
  {O'Sullivan}, \citenamefont {Mirhosseini}, \citenamefont {Robertson},
  \citenamefont {Padgett},\ and\ \citenamefont {Boyd}}]{BRodenburg2012a}%
  \BibitemOpen
  \bibfield  {author} {\bibinfo {author} {\bibfnamefont {B.}~\bibnamefont
  {Rodenburg}}, \bibinfo {author} {\bibfnamefont {M.~P.~J.}\ \bibnamefont
  {Lavery}}, \bibinfo {author} {\bibfnamefont {M.}~\bibnamefont {Malik}},
  \bibinfo {author} {\bibfnamefont {M.~N.}\ \bibnamefont {O'Sullivan}},
  \bibinfo {author} {\bibfnamefont {M.}~\bibnamefont {Mirhosseini}}, \bibinfo
  {author} {\bibfnamefont {D.~J.}\ \bibnamefont {Robertson}}, \bibinfo {author}
  {\bibfnamefont {M.}~\bibnamefont {Padgett}}, \ and\ \bibinfo {author}
  {\bibfnamefont {R.~W.}\ \bibnamefont {Boyd}},\ }\href@noop {} {\bibfield
  {journal} {\bibinfo  {journal} {Opt. Lett.}\ }\textbf {\bibinfo {volume}
  {37}},\ \bibinfo {pages} {3735} (\bibinfo {year} {2012})}\BibitemShut
  {NoStop}%
\bibitem [{\citenamefont {Rodenburg}\ \emph
  {et~al.}(2014{\natexlab{a}})\citenamefont {Rodenburg}, \citenamefont
  {Mirhosseini}, \citenamefont {Malik}, \citenamefont {Maga{\~n}a-Loaiza},
  \citenamefont {Yanakas} \emph {et~al.}}]{rodenburg2014a}%
  \BibitemOpen
  \bibfield  {author} {\bibinfo {author} {\bibfnamefont {B.}~\bibnamefont
  {Rodenburg}}, \bibinfo {author} {\bibfnamefont {M.}~\bibnamefont
  {Mirhosseini}}, \bibinfo {author} {\bibfnamefont {M.}~\bibnamefont {Malik}},
  \bibinfo {author} {\bibfnamefont {O.~S.}\ \bibnamefont {Maga{\~n}a-Loaiza}},
  \bibinfo {author} {\bibfnamefont {M.}~\bibnamefont {Yanakas}},  \emph
  {et~al.},\ }\href@noop {} {\bibfield  {journal} {\bibinfo  {journal} {New J.
  Phys.}\ }\textbf {\bibinfo {volume} {16}},\ \bibinfo {pages} {033020}
  (\bibinfo {year} {2014}{\natexlab{a}})}\BibitemShut {NoStop}%
\bibitem [{\citenamefont {Rodenburg}\ \emph
  {et~al.}(2014{\natexlab{b}})\citenamefont {Rodenburg}, \citenamefont
  {Mirhosseini}, \citenamefont {Malik}, \citenamefont {Maga{\~n}a-Loaiza},
  \citenamefont {Yanakas} \emph {et~al.}}]{rodenburg2014err}%
  \BibitemOpen
  \bibfield  {author} {\bibinfo {author} {\bibfnamefont {B.}~\bibnamefont
  {Rodenburg}}, \bibinfo {author} {\bibfnamefont {M.}~\bibnamefont
  {Mirhosseini}}, \bibinfo {author} {\bibfnamefont {M.}~\bibnamefont {Malik}},
  \bibinfo {author} {\bibfnamefont {O.~S.}\ \bibnamefont {Maga{\~n}a-Loaiza}},
  \bibinfo {author} {\bibfnamefont {M.}~\bibnamefont {Yanakas}},  \emph
  {et~al.},\ }\href@noop {} {\bibfield  {journal} {\bibinfo  {journal} {New J.
  Phys.}\ }\textbf {\bibinfo {volume} {16}},\ \bibinfo {pages} {089501(E)}
  (\bibinfo {year} {2014}{\natexlab{b}})}\BibitemShut {NoStop}%
\bibitem [{\citenamefont {Paterson}(2005)}]{Paterson2005}%
  \BibitemOpen
  \bibfield  {author} {\bibinfo {author} {\bibfnamefont {C.}~\bibnamefont
  {Paterson}},\ }\href@noop {} {\bibfield  {journal} {\bibinfo  {journal}
  {Phys. Rev. Lett.}\ }\textbf {\bibinfo {volume} {94}},\ \bibinfo {pages}
  {153901} (\bibinfo {year} {2005})}\BibitemShut {NoStop}%
\bibitem [{\citenamefont {Gapaul}\ and\ \citenamefont
  {Andrews}(2007)}]{Gopaul2007}%
  \BibitemOpen
  \bibfield  {author} {\bibinfo {author} {\bibfnamefont {C.}~\bibnamefont
  {Gapaul}}\ and\ \bibinfo {author} {\bibfnamefont {R.}~\bibnamefont
  {Andrews}},\ }\href@noop {} {\bibfield  {journal} {\bibinfo  {journal} {New
  J. Phys.}\ }\textbf {\bibinfo {volume} {9}},\ \bibinfo {pages} {94} (\bibinfo
  {year} {2007})}\BibitemShut {NoStop}%
\bibitem [{\citenamefont {Tyler}\ and\ \citenamefont {Boyd}(2009)}]{Tyler2009}%
  \BibitemOpen
  \bibfield  {author} {\bibinfo {author} {\bibfnamefont {G.~A.}\ \bibnamefont
  {Tyler}}\ and\ \bibinfo {author} {\bibfnamefont {R.~W.}\ \bibnamefont
  {Boyd}},\ }\href@noop {} {\bibfield  {journal} {\bibinfo  {journal} {Opt.
  Lett.}\ }\textbf {\bibinfo {volume} {34}},\ \bibinfo {pages} {142} (\bibinfo
  {year} {2009})}\BibitemShut {NoStop}%
\bibitem [{\citenamefont {Anguita}\ \emph {et~al.}(2008)\citenamefont
  {Anguita}, \citenamefont {Neifeld},\ and\ \citenamefont
  {Vasic}}]{Anguita2008}%
  \BibitemOpen
  \bibfield  {author} {\bibinfo {author} {\bibfnamefont {J.~A.}\ \bibnamefont
  {Anguita}}, \bibinfo {author} {\bibfnamefont {M.~A.}\ \bibnamefont
  {Neifeld}}, \ and\ \bibinfo {author} {\bibfnamefont {B.~V.}\ \bibnamefont
  {Vasic}},\ }\href@noop {} {\bibfield  {journal} {\bibinfo  {journal} {Appl.
  Opt.}\ }\textbf {\bibinfo {volume} {47}},\ \bibinfo {pages} {2414} (\bibinfo
  {year} {2008})}\BibitemShut {NoStop}%
\bibitem [{\citenamefont {Smith}\ and\ \citenamefont
  {Raymer}(2006)}]{Smith2006}%
  \BibitemOpen
  \bibfield  {author} {\bibinfo {author} {\bibfnamefont {B.~J.}\ \bibnamefont
  {Smith}}\ and\ \bibinfo {author} {\bibfnamefont {M.~G.}\ \bibnamefont
  {Raymer}},\ }\href@noop {} {\bibfield  {journal} {\bibinfo  {journal} {Phys.
  Rev. A}\ }\textbf {\bibinfo {volume} {74}},\ \bibinfo {pages} {062104}
  (\bibinfo {year} {2006})}\BibitemShut {NoStop}%
\bibitem [{\citenamefont {Roux}(2011)}]{Roux2011}%
  \BibitemOpen
  \bibfield  {author} {\bibinfo {author} {\bibfnamefont {F.~S.}\ \bibnamefont
  {Roux}},\ }\href@noop {} {\bibfield  {journal} {\bibinfo  {journal} {Phys.
  Rev. A}\ }\textbf {\bibinfo {volume} {83}},\ \bibinfo {pages} {053822}
  (\bibinfo {year} {2011})}\BibitemShut {NoStop}%
\bibitem [{\citenamefont {Zhang}\ \emph {et~al.}(2012)\citenamefont {Zhang},
  \citenamefont {Djordjevic},\ and\ \citenamefont {Gao}}]{zhang2012}%
  \BibitemOpen
  \bibfield  {author} {\bibinfo {author} {\bibfnamefont {Y.}~\bibnamefont
  {Zhang}}, \bibinfo {author} {\bibfnamefont {I.~B.}\ \bibnamefont
  {Djordjevic}}, \ and\ \bibinfo {author} {\bibfnamefont {X.}~\bibnamefont
  {Gao}},\ }\href@noop {} {\bibfield  {journal} {\bibinfo  {journal} {Opt.
  Lett.}\ }\textbf {\bibinfo {volume} {37}},\ \bibinfo {pages} {3267} (\bibinfo
  {year} {2012})}\BibitemShut {NoStop}%
\bibitem [{\citenamefont {Pors}\ \emph {et~al.}(2011)\citenamefont {Pors},
  \citenamefont {Monken}, \citenamefont {Eliel},\ and\ \citenamefont
  {Woerdman}}]{pors}%
  \BibitemOpen
  \bibfield  {author} {\bibinfo {author} {\bibfnamefont {B.-J.}\ \bibnamefont
  {Pors}}, \bibinfo {author} {\bibfnamefont {C.~H.}\ \bibnamefont {Monken}},
  \bibinfo {author} {\bibfnamefont {E.~R.}\ \bibnamefont {Eliel}}, \ and\
  \bibinfo {author} {\bibfnamefont {J.~P.}\ \bibnamefont {Woerdman}},\
  }\href@noop {} {\bibfield  {journal} {\bibinfo  {journal} {Opt. Express}\
  }\textbf {\bibinfo {volume} {19}},\ \bibinfo {pages} {6671} (\bibinfo {year}
  {2011})}\BibitemShut {NoStop}%
\bibitem [{\citenamefont {Hamadou~Ibrahim}\ \emph {et~al.}(2013)\citenamefont
  {Hamadou~Ibrahim}, \citenamefont {Roux}, \citenamefont {McLaren},
  \citenamefont {Konrad},\ and\ \citenamefont {Forbes}}]{Ibrahim2013}%
  \BibitemOpen
  \bibfield  {author} {\bibinfo {author} {\bibfnamefont {A.}~\bibnamefont
  {Hamadou~Ibrahim}}, \bibinfo {author} {\bibfnamefont {F.~S.}\ \bibnamefont
  {Roux}}, \bibinfo {author} {\bibfnamefont {M.}~\bibnamefont {McLaren}},
  \bibinfo {author} {\bibfnamefont {T.}~\bibnamefont {Konrad}}, \ and\ \bibinfo
  {author} {\bibfnamefont {A.}~\bibnamefont {Forbes}},\ }\href@noop {}
  {\bibfield  {journal} {\bibinfo  {journal} {Phys. Rev. A}\ }\textbf {\bibinfo
  {volume} {88}},\ \bibinfo {pages} {012312} (\bibinfo {year}
  {2013})}\BibitemShut {NoStop}%
\bibitem [{\citenamefont {Zhao}\ \emph {et~al.}(2012)\citenamefont {Zhao},
  \citenamefont {Leach}, \citenamefont {Gong}, \citenamefont {Ding},\ and\
  \citenamefont {Zheng}}]{leachao}%
  \BibitemOpen
  \bibfield  {author} {\bibinfo {author} {\bibfnamefont {S.~M.}\ \bibnamefont
  {Zhao}}, \bibinfo {author} {\bibfnamefont {J.}~\bibnamefont {Leach}},
  \bibinfo {author} {\bibfnamefont {L.~Y.}\ \bibnamefont {Gong}}, \bibinfo
  {author} {\bibfnamefont {J.}~\bibnamefont {Ding}}, \ and\ \bibinfo {author}
  {\bibfnamefont {B.~Y.}\ \bibnamefont {Zheng}},\ }\href@noop {} {\bibfield
  {journal} {\bibinfo  {journal} {Opt. Express}\ }\textbf {\bibinfo {volume}
  {20}},\ \bibinfo {pages} {452} (\bibinfo {year} {2012})}\BibitemShut
  {NoStop}%
\bibitem [{\citenamefont {Ren}\ \emph {et~al.}(2014)\citenamefont {Ren},
  \citenamefont {Xie}, \citenamefont {Huang}, \citenamefont {Ahmed} \emph
  {et~al.}}]{renetal}%
  \BibitemOpen
  \bibfield  {author} {\bibinfo {author} {\bibfnamefont {Y.}~\bibnamefont
  {Ren}}, \bibinfo {author} {\bibfnamefont {G.}~\bibnamefont {Xie}}, \bibinfo
  {author} {\bibfnamefont {H.}~\bibnamefont {Huang}}, \bibinfo {author}
  {\bibfnamefont {N.}~\bibnamefont {Ahmed}},  \emph {et~al.},\ }\href@noop {}
  {\bibfield  {journal} {\bibinfo  {journal} {Optica}\ }\textbf {\bibinfo
  {volume} {1}},\ \bibinfo {pages} {376} (\bibinfo {year} {2014})}\BibitemShut
  {NoStop}%
\bibitem [{\citenamefont {Br\"unner}\ and\ \citenamefont
  {Roux}(2013)}]{robust}%
  \BibitemOpen
  \bibfield  {author} {\bibinfo {author} {\bibfnamefont {T.}~\bibnamefont
  {Br\"unner}}\ and\ \bibinfo {author} {\bibfnamefont {F.~S.}\ \bibnamefont
  {Roux}},\ }\href@noop {} {\bibfield  {journal} {\bibinfo  {journal} {New J.
  Phys.}\ }\textbf {\bibinfo {volume} {15}},\ \bibinfo {pages} {063005}
  (\bibinfo {year} {2013})}\BibitemShut {NoStop}%
\bibitem [{\citenamefont {Gonzalez~Alonso}\ and\ \citenamefont
  {Brun}(2013)}]{brun}%
  \BibitemOpen
  \bibfield  {author} {\bibinfo {author} {\bibfnamefont {J.~R.}\ \bibnamefont
  {Gonzalez~Alonso}}\ and\ \bibinfo {author} {\bibfnamefont {T.~A.}\
  \bibnamefont {Brun}},\ }\href@noop {} {\bibfield  {journal} {\bibinfo
  {journal} {Phys. Rev. A}\ }\textbf {\bibinfo {volume} {88}},\ \bibinfo
  {pages} {022326} (\bibinfo {year} {2013})}\BibitemShut {NoStop}%
\bibitem [{\citenamefont {Mafu}\ \emph {et~al.}(2013)\citenamefont {Mafu},
  \citenamefont {Dudley}, \citenamefont {Goyal}, \citenamefont {Giovannini},
  \citenamefont {McLaren}, \citenamefont {Padgett}, \citenamefont {Konrad},
  \citenamefont {Petruccione}, \citenamefont {L\"utkenhaus},\ and\
  \citenamefont {Forbes}}]{Mafu2013}%
  \BibitemOpen
  \bibfield  {author} {\bibinfo {author} {\bibfnamefont {M.}~\bibnamefont
  {Mafu}}, \bibinfo {author} {\bibfnamefont {A.}~\bibnamefont {Dudley}},
  \bibinfo {author} {\bibfnamefont {S.}~\bibnamefont {Goyal}}, \bibinfo
  {author} {\bibfnamefont {D.}~\bibnamefont {Giovannini}}, \bibinfo {author}
  {\bibfnamefont {M.}~\bibnamefont {McLaren}}, \bibinfo {author} {\bibfnamefont
  {M.~J.}\ \bibnamefont {Padgett}}, \bibinfo {author} {\bibfnamefont
  {T.}~\bibnamefont {Konrad}}, \bibinfo {author} {\bibfnamefont
  {F.}~\bibnamefont {Petruccione}}, \bibinfo {author} {\bibfnamefont
  {N.}~\bibnamefont {L\"utkenhaus}}, \ and\ \bibinfo {author} {\bibfnamefont
  {A.}~\bibnamefont {Forbes}},\ }\href@noop {} {\bibfield  {journal} {\bibinfo
  {journal} {Phys. Rev. A}\ }\textbf {\bibinfo {volume} {88}},\ \bibinfo
  {pages} {032305} (\bibinfo {year} {2013})}\BibitemShut {NoStop}%
\bibitem [{\citenamefont {Vallone}\ \emph {et~al.}(2014)\citenamefont
  {Vallone}, \citenamefont {D'Ambrosio}, \citenamefont {Sponselli},
  \citenamefont {Slussarenko}, \citenamefont {Marrucci}, \citenamefont
  {Sciarrino},\ and\ \citenamefont {Villoresi}}]{vallone2014}%
  \BibitemOpen
  \bibfield  {author} {\bibinfo {author} {\bibfnamefont {G.}~\bibnamefont
  {Vallone}}, \bibinfo {author} {\bibfnamefont {V.}~\bibnamefont {D'Ambrosio}},
  \bibinfo {author} {\bibfnamefont {A.}~\bibnamefont {Sponselli}}, \bibinfo
  {author} {\bibfnamefont {S.}~\bibnamefont {Slussarenko}}, \bibinfo {author}
  {\bibfnamefont {L.}~\bibnamefont {Marrucci}}, \bibinfo {author}
  {\bibfnamefont {F.}~\bibnamefont {Sciarrino}}, \ and\ \bibinfo {author}
  {\bibfnamefont {P.}~\bibnamefont {Villoresi}},\ }\href@noop {} {\bibfield
  {journal} {\bibinfo  {journal} {Phys. Rev. Lett.}\ }\textbf {\bibinfo
  {volume} {113}},\ \bibinfo {pages} {060503} (\bibinfo {year}
  {2014})}\BibitemShut {NoStop}%
\bibitem [{\citenamefont {Mirhosseini}\ \emph {et~al.}(2014)\citenamefont
  {Mirhosseini}, \citenamefont {Maga{\~n}a-Loaiza}, \citenamefont {O'Sullivan},
  \citenamefont {Rodenburg}, \citenamefont {Malik}, \citenamefont {Gauthier},\
  and\ \citenamefont {Boyd}}]{mir2014}%
  \BibitemOpen
  \bibfield  {author} {\bibinfo {author} {\bibfnamefont {M.}~\bibnamefont
  {Mirhosseini}}, \bibinfo {author} {\bibfnamefont {O.~S.}\ \bibnamefont
  {Maga{\~n}a-Loaiza}}, \bibinfo {author} {\bibfnamefont {M.~N.}\ \bibnamefont
  {O'Sullivan}}, \bibinfo {author} {\bibfnamefont {B.}~\bibnamefont
  {Rodenburg}}, \bibinfo {author} {\bibfnamefont {M.}~\bibnamefont {Malik}},
  \bibinfo {author} {\bibfnamefont {D.~J.}\ \bibnamefont {Gauthier}}, \ and\
  \bibinfo {author} {\bibfnamefont {R.~W.}\ \bibnamefont {Boyd}},\ }\href@noop
  {} {\bibfield  {journal} {\bibinfo  {journal} {arXiv preprint
  arXiv:1402.7113}\ } (\bibinfo {year} {2014})}\BibitemShut {NoStop}%
\bibitem [{\citenamefont {Ekert}(1991)}]{Ekert1991}%
  \BibitemOpen
  \bibfield  {author} {\bibinfo {author} {\bibfnamefont {A.}~\bibnamefont
  {Ekert}},\ }\href@noop {} {\bibfield  {journal} {\bibinfo  {journal} {Phys.
  Rev. Lett.}\ }\textbf {\bibinfo {volume} {67}},\ \bibinfo {pages} {661}
  (\bibinfo {year} {1991})}\BibitemShut {NoStop}%
\bibitem [{\citenamefont {Bru\ss}(1998)}]{Bruss1998}%
  \BibitemOpen
  \bibfield  {author} {\bibinfo {author} {\bibfnamefont {D.}~\bibnamefont
  {Bru\ss}},\ }\href@noop {} {\bibfield  {journal} {\bibinfo  {journal} {Phys.
  Rev. Lett.}\ }\textbf {\bibinfo {volume} {81}},\ \bibinfo {pages} {3018}
  (\bibinfo {year} {1998})}\BibitemShut {NoStop}%
\bibitem [{\citenamefont {Schwinger}(1960)}]{schwinger1960}%
  \BibitemOpen
  \bibfield  {author} {\bibinfo {author} {\bibfnamefont {J.}~\bibnamefont
  {Schwinger}},\ }\href@noop {} {\bibfield  {journal} {\bibinfo  {journal}
  {Proceedings of the national academy of sciences of the United States Of
  America}\ }\textbf {\bibinfo {volume} {46}},\ \bibinfo {pages} {570}
  (\bibinfo {year} {1960})}\BibitemShut {NoStop}%
\bibitem [{\citenamefont {Bennett}\ \emph {et~al.}(1984)\citenamefont
  {Bennett}, \citenamefont {Brassard} \emph {et~al.}}]{Bennett1984}%
  \BibitemOpen
  \bibfield  {author} {\bibinfo {author} {\bibfnamefont {C.}~\bibnamefont
  {Bennett}}, \bibinfo {author} {\bibfnamefont {G.}~\bibnamefont {Brassard}},
  \emph {et~al.},\ }in\ \href@noop {} {\emph {\bibinfo {booktitle} {Proceedings
  of IEEE International Conference on Computers, Systems and Signal
  Processing}}},\ Vol.\ \bibinfo {volume} {175}\ (\bibinfo {organization}
  {Bangalore, India},\ \bibinfo {year} {1984})\BibitemShut {NoStop}%
\bibitem [{\citenamefont {Ferenczi}\ and\ \citenamefont
  {L\"utkenhaus}(2012)}]{Ferenczi2012}%
  \BibitemOpen
  \bibfield  {author} {\bibinfo {author} {\bibfnamefont {A.}~\bibnamefont
  {Ferenczi}}\ and\ \bibinfo {author} {\bibfnamefont {N.}~\bibnamefont
  {L\"utkenhaus}},\ }\href@noop {} {\bibfield  {journal} {\bibinfo  {journal}
  {Phys. Rev. A}\ }\textbf {\bibinfo {volume} {85}},\ \bibinfo {pages} {052310}
  (\bibinfo {year} {2012})}\BibitemShut {NoStop}%
\bibitem [{\citenamefont {Sheridan}\ and\ \citenamefont
  {Scarani}(2010)}]{Sheridan2010}%
  \BibitemOpen
  \bibfield  {author} {\bibinfo {author} {\bibfnamefont {L.}~\bibnamefont
  {Sheridan}}\ and\ \bibinfo {author} {\bibfnamefont {V.}~\bibnamefont
  {Scarani}},\ }\href@noop {} {\bibfield  {journal} {\bibinfo  {journal} {Phys.
  Rev. A}\ }\textbf {\bibinfo {volume} {82}},\ \bibinfo {pages} {030301}
  (\bibinfo {year} {2010})}\BibitemShut {NoStop}%
\bibitem [{\citenamefont {Horodecki}\ \emph {et~al.}(2009)\citenamefont
  {Horodecki}, \citenamefont {Horodecki}, \citenamefont {Horodecki},\ and\
  \citenamefont {Horodecki}}]{Horodecki-review}%
  \BibitemOpen
  \bibfield  {author} {\bibinfo {author} {\bibfnamefont {R.}~\bibnamefont
  {Horodecki}}, \bibinfo {author} {\bibfnamefont {P.}~\bibnamefont
  {Horodecki}}, \bibinfo {author} {\bibfnamefont {M.}~\bibnamefont
  {Horodecki}}, \ and\ \bibinfo {author} {\bibfnamefont {K.}~\bibnamefont
  {Horodecki}},\ }\href@noop {} {\bibfield  {journal} {\bibinfo  {journal}
  {Rev. Mod. Phys.}\ }\textbf {\bibinfo {volume} {81}},\ \bibinfo {pages} {865}
  (\bibinfo {year} {2009})}\BibitemShut {NoStop}%
\bibitem [{\citenamefont {Wootters}(1998)}]{Wootters1998}%
  \BibitemOpen
  \bibfield  {author} {\bibinfo {author} {\bibfnamefont {W.~K.}\ \bibnamefont
  {Wootters}},\ }\href@noop {} {\bibfield  {journal} {\bibinfo  {journal}
  {Phys. Rev. Lett.}\ }\textbf {\bibinfo {volume} {80}},\ \bibinfo {pages}
  {2245} (\bibinfo {year} {1998})}\BibitemShut {NoStop}%
\bibitem [{\citenamefont {Kolmogorov}(1941)}]{kolmogorov}%
  \BibitemOpen
  \bibfield  {author} {\bibinfo {author} {\bibfnamefont {A.~N.}\ \bibnamefont
  {Kolmogorov}},\ }\href@noop {} {\bibfield  {journal} {\bibinfo  {journal}
  {Akademiia Nauk SSSR Doklady}\ }\textbf {\bibinfo {volume} {30}},\ \bibinfo
  {pages} {301} (\bibinfo {year} {1941})}\BibitemShut {NoStop}%
\bibitem [{\citenamefont {Martin}\ and\ \citenamefont {Flatt\'e}(1988)}]{MF1}%
  \BibitemOpen
  \bibfield  {author} {\bibinfo {author} {\bibfnamefont {J.~M.}\ \bibnamefont
  {Martin}}\ and\ \bibinfo {author} {\bibfnamefont {S.~M.}\ \bibnamefont
  {Flatt\'e}},\ }\href@noop {} {\bibfield  {journal} {\bibinfo  {journal}
  {Appl. Opt.}\ }\textbf {\bibinfo {volume} {27}},\ \bibinfo {pages} {2111}
  (\bibinfo {year} {1988})}\BibitemShut {NoStop}%
\bibitem [{\citenamefont {Martin}\ and\ \citenamefont {Flatt\'e}(1990)}]{MF2}%
  \BibitemOpen
  \bibfield  {author} {\bibinfo {author} {\bibfnamefont {J.~M.}\ \bibnamefont
  {Martin}}\ and\ \bibinfo {author} {\bibfnamefont {S.~M.}\ \bibnamefont
  {Flatt\'e}},\ }\href@noop {} {\bibfield  {journal} {\bibinfo  {journal} {J.
  Opt. Soc. Am. A}\ }\textbf {\bibinfo {volume} {7}},\ \bibinfo {pages} {838}
  (\bibinfo {year} {1990})}\BibitemShut {NoStop}%
\bibitem [{\citenamefont {Knepp}(1983)}]{KNEPP1983}%
  \BibitemOpen
  \bibfield  {author} {\bibinfo {author} {\bibfnamefont {D.~L.}\ \bibnamefont
  {Knepp}},\ }\href@noop {} {\bibfield  {journal} {\bibinfo  {journal} {Proc.
  IEEE}\ }\textbf {\bibinfo {volume} {71}},\ \bibinfo {pages} {722} (\bibinfo
  {year} {1983})}\BibitemShut {NoStop}%
\bibitem [{\citenamefont {Lane}\ \emph {et~al.}(1992)\citenamefont {Lane},
  \citenamefont {Glindemann},\ and\ \citenamefont {Dainty}}]{Lane1992}%
  \BibitemOpen
  \bibfield  {author} {\bibinfo {author} {\bibfnamefont {R.~G.}\ \bibnamefont
  {Lane}}, \bibinfo {author} {\bibfnamefont {A.}~\bibnamefont {Glindemann}}, \
  and\ \bibinfo {author} {\bibfnamefont {J.~C.}\ \bibnamefont {Dainty}},\
  }\href@noop {} {\bibfield  {journal} {\bibinfo  {journal} {Waves in Random
  Media}\ }\textbf {\bibinfo {volume} {2}},\ \bibinfo {pages} {209} (\bibinfo
  {year} {1992})}\BibitemShut {NoStop}%
\end{thebibliography}

%

\end{document}